\documentclass[a4paper,11pt]{article}
\usepackage{pos}
\usepackage{slashed}
\usepackage{amsmath}
\usepackage{graphicx}
\title{Searching for top squarks from the landscape at HL-LHC}
\author*[a]{Juhi Dutta}
 \affiliation[a]{Homer L. Dodge Department of Physics and Astronomy,
University of Oklahoma, Norman, OK 73019, USA}
  \emailAdd{juhi.dutta@ou.edu}

\abstract{Supersymmetric models with low electroweak fine-tuning are more prevalent on the string landscape than fine-tuned models. We assume a fertile patch of landscape vacua containing the minimal supersymmetric standard model (MSSM) as a low-energy EFT. Such models are characterized by light higgsinos in the mass range of a few hundred GeV whilst top squarks are in the 1-2.5 TeV range. Other sparticles are generally beyond current LHC reach. We evaluate prospects for top squark searches of the expected natural SUSY at HL-LHC.}

\FullConference{The European Physical Society Conference on High Energy Physics (EPS-HEP2023)\\
 21-25 August 2023\\
Hamburg, Germany\\}

\begin{document}
\maketitle
\section{Introduction}
Supersymmetry is one of the best-motivated candidates for Beyond Standard Model (BSM) to address several issues of Nature. While early estimates of naturalness  placed stringent upper limits on masses of supersymmetric particles, such as  $m_{\widetilde{t}_1} \leq$ 300-400 GeV for $\Delta_{BG} < $ 10-30  where $\Delta_{BG}$ is the Barbieri-Guidice measure of naturalness~\cite{Barbieri:1987fn}, defined as   
\begin{equation*}
    \Delta_{BG}=max_i|\frac{p_i\partial m^2_Z}{m^2_Z\partial dp_i}|.
\end{equation*}  
 where $p_i$ are the fundamental free parameters of the theory
and subsequent measures such as $\Delta_{HS}$ required third generation squarks to be less than 500 GeV\cite{Kitano:2006gv, Brust:2011tb},   current experimental searches from LHC place rather stringent limits $\sim \mathcal{O}$(TeV) on the lightest stops.   However, the naturalness measures considered earlier turned out to be large overestimates of the actual-finetuning\cite{Baer:2013gva,Mustafayev:2014lqa,Baer:2014ica,Baer:2023cvi}. 
Recently, more conservative estimates of electroweak fine-tuning $\Delta_{EW}$~\cite{Baer:2012up} have come up to resolve the naturalness issue where $\Delta_{EW}$ is
\begin{equation}
     \frac{m^{2}_Z}{2}=\frac{m^2_{H_d}+\Sigma^d_d-(m^2_{H_u}+\Sigma^u_u)\tan^2 \beta}{\tan^2 \beta -1 }-\mu^2
     \label{eq:mz}
\end{equation}
the ratio of the largest term on the right-hand side of the eq.~\ref{eq:mz} to $\frac{m^2_Z}{2}$, where $m_Z$ is the mass of the $Z$ boson, $m^2_{H_u}$ and $m^2_{H_d}$ refer to the Higgs soft breaking masses coupling to the up-type and down-type quarks respectively, $\tan \beta = \frac{v_u}{v_d}$ ($v_q$ being the vaccuum expectation value of $H_q$, where $q=(u,d)$), and $\Sigma^q_q$, refer to the loop contributions from the particles and sparticles to the Higgs sector (with dominant one-loop contributions from the lightest top squarks) and $\mu$ is the higgsino mass parameter. Such a conservative measure of naturalness imposes relatively relaxed constraints on sparticle masses which are allowed up to several TeV at little cost to finetuning since their contributions to the weak scale are suppressed by loop factors.%

Another possible resolution arises from the string landscape picture.  In the string landscape, where order of 10$^{500}$ vacua solutions arise from compactification from 10 to 4 spacetime dimensions, each vacuum solution corresponds to a different set of 4-d low energy effective field theory law of physics.  The string landscape provides a natural setting for Weinberg’s anthropic solution to the cosmological constant problem~\cite{Weinberg:1987dv} in an eternally inflating multiverse. In the same spirit, one tries to address the origin of the SUSY breaking scale in the string landscape. Supersymmetric models with low electroweak fine-tuning are expected to be more prevalent on the string landscape than fine-tuned models~\cite{Baer:2020kwz}. In this work, we determine the properties of the stops from the landscape and prospects of observing them at the upcoming HL-LHC.  
\section{MSSM from the  string landscape}
Assuming a fertile patch of landscape vacua containing the minimal supersymmetric standard model (MSSM) as low energy effective field theory~\cite{Baer:2023cvi}, the landscape statistically favours large soft terms via a power law~\cite{Douglas:2012bu},
\begin{equation*}
    f_{SUSY} = m^{2n_F+n_D-1}_{soft}
\end{equation*}
where $n_F$ and $n_D$ refer to the number of F-term and D-term SUSY breaking terms and where $f_{\text{SUSY}}$ is the expected statistical distribution of landscape soft terms. 
A statistical pull by the landscape to large soft terms is balanced by the requirement of a derived value of the weak scale in the pocket universe ($PU$), which is not too far from its measured value in our universe ($OU$) given by the ABDS window~\cite{Agrawal:1997gf}
\begin{equation*}
    \frac{m^{PU2}_Z}{2}=\frac{m^2_{H_d}+\Sigma^d_d-(m^2_{H_u}+\Sigma^u_u)\tan^2 \beta}{\tan^2 \beta -1 }-\mu^2_{PU}
\end{equation*}
such that $m^{PU}_{weak}\sim ( 0.5-5 ) m^{OU}_{weak}$ in order to allow for complex nuclei (and hence atoms) in our universe.  
 The string landscape approach to soft SUSY breaking within the
MSSM statistically predicts a Higgs boson with mass 125 GeV and is characterized by light higgsinos in the 100-400 GeV range,  lightest top squarks are in the 1-2.5 TeV range with large trilinear soft terms which helps to push $m_h$ $\sim$ 125 GeV and other squarks beyond the HL-LHC reach.
\begin{figure}
    \centering
    \includegraphics[scale=0.14]{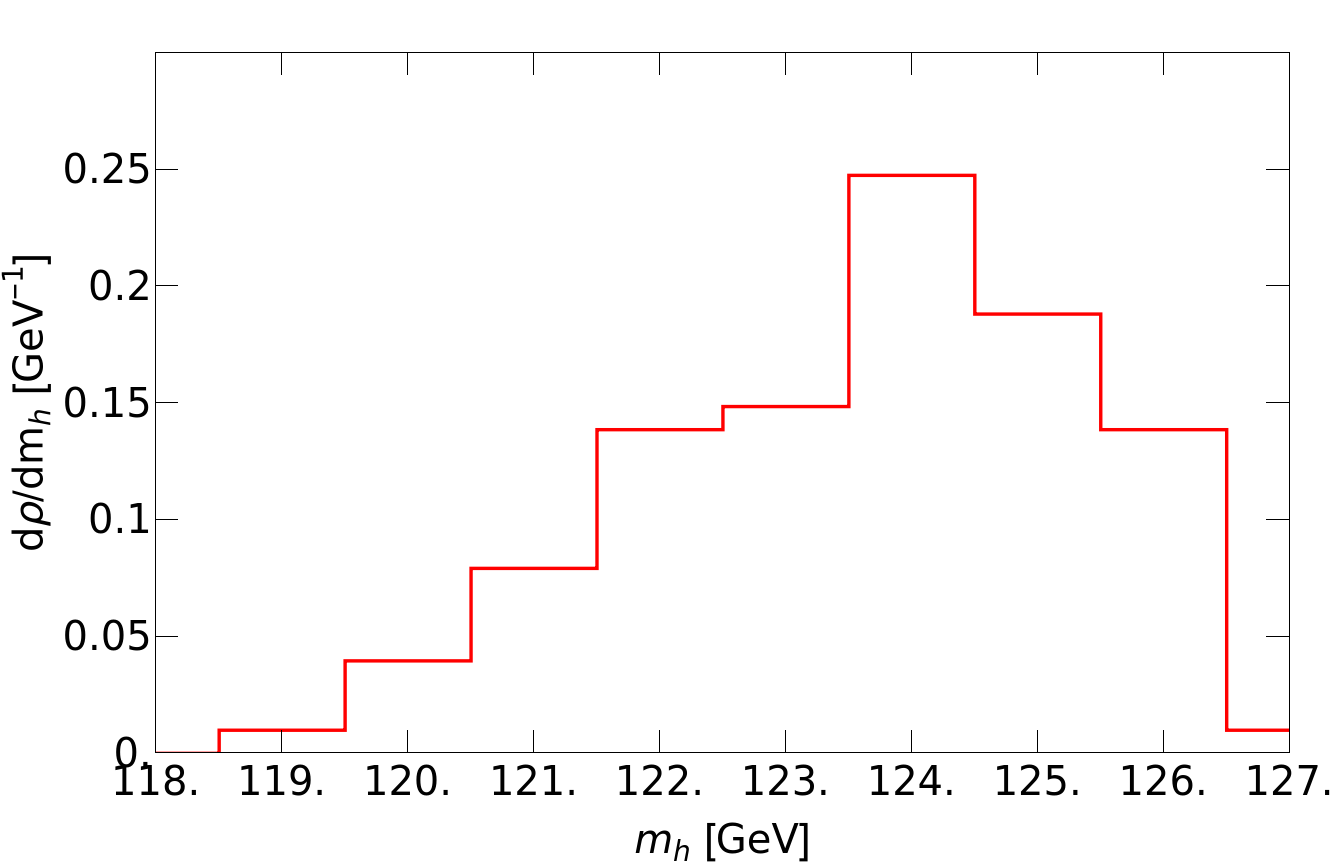}
    \caption{The probability distributions for the lightest CP-even Higgs mass $m_h$. }
    \label{fig:mh}
\end{figure}
To obtain a measure of \textit{stringy naturalness}, we implement a linear scan\cite{Baer:2023uwo} over the NUHM2 parameter space. Fig.~\ref{fig:mh} shows the probability distribution for the lightest CP-even Higgs mass, $m_h$ from the string landscape with a n=1 draw to large soft terms. We observe that the distribution is peaked towards $\sim$125 GeV while $m_{\widetilde{t}_1}$ has a large number of events above the TeV scale within 1-2.5 TeV with a peak $\sim1.5$ TeV.  The reach of LHC Run 3 and
HL-LHC would probe the  peak probability region in the coming years, making the search for light top-squarks of supersymmetry a highly motivated priority. 
\begin{figure}
    \centering
      \includegraphics[scale=0.14]{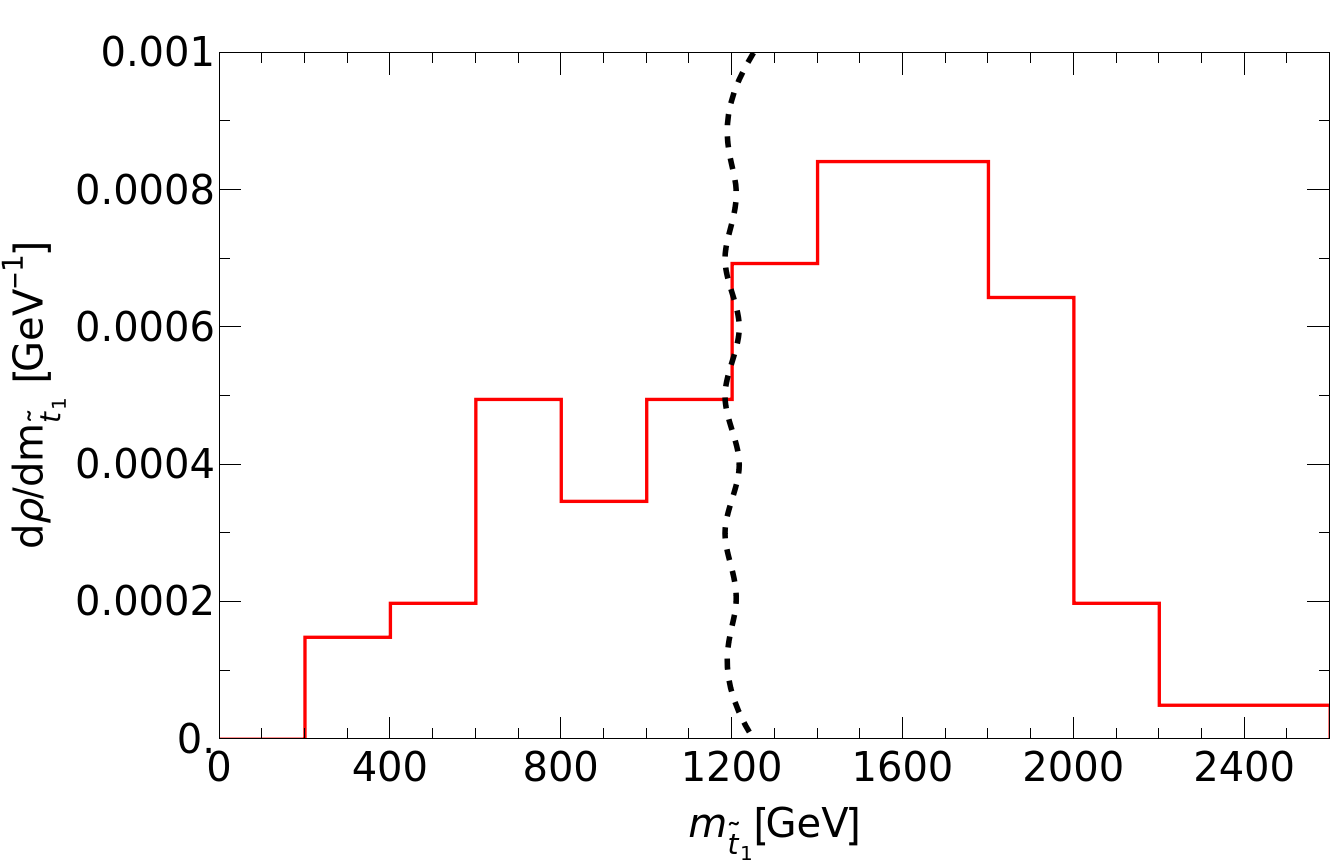}
      \includegraphics[scale=0.14]{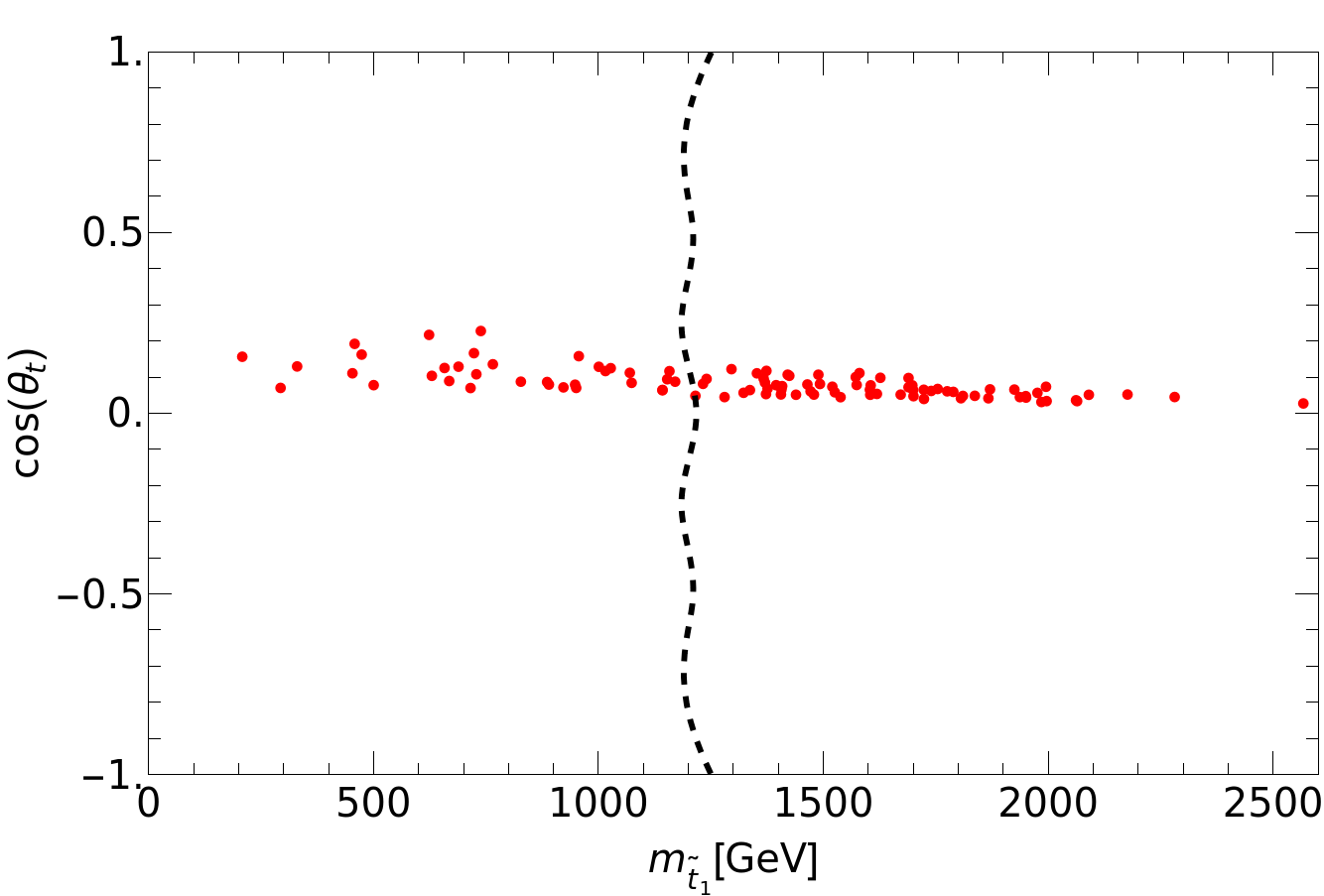}
    \caption{The probability distribution of the lightest top squark $m_{\widetilde{t}_1}$ (left) and mixing angle $\cos \theta$ in the stop sector (right). }
    \label{fig:stops}
\end{figure} 
Fig.~\ref{fig:stops} (right) shows the variation of the cosine of mixing angle in the stop sector vs. the lightest stop mass. It is clear from the plot, $\cos \theta \sim 0.1$ over most of the stop mass range suggesting that the lightest stop is largely dominated by the right-handed top squark. Thus, the light top-squark decays comparably via  $b\widetilde{\chi}^{\pm}_1$ and $t\widetilde{\chi}^0_i$ yielding mixed final states of $b\bar{b}+ \slashed{E}_T$ , $t\bar{b}/ t\bar{b} + \slashed{E}_T$ and $t\bar{t}+ \slashed{E}_T$.
\label{subsec:prod}      
\begin{figure}[ht]
    \centering
    \includegraphics[scale=0.135]{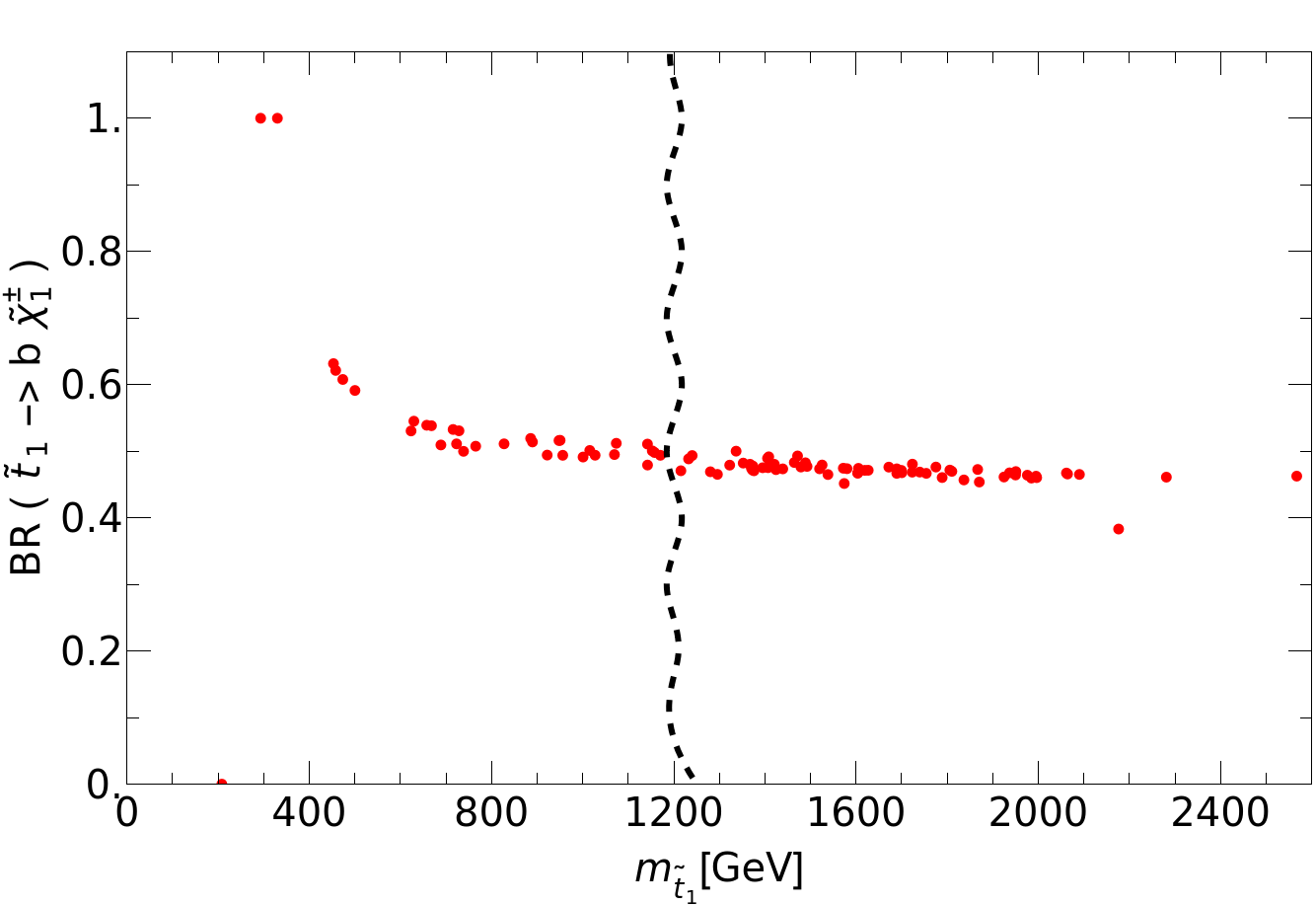}
    \includegraphics[scale=0.135]{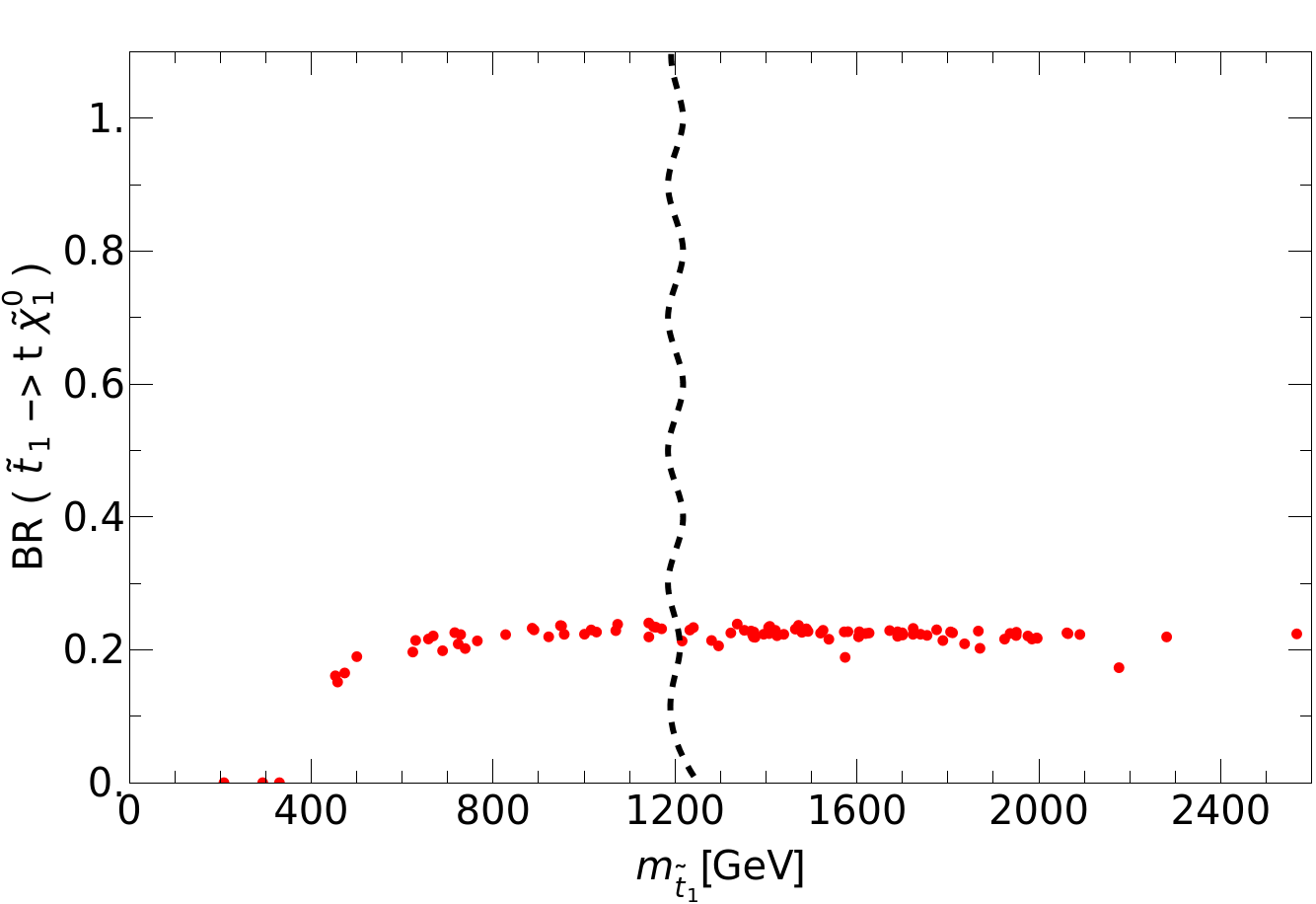}
    \includegraphics[scale=0.135]{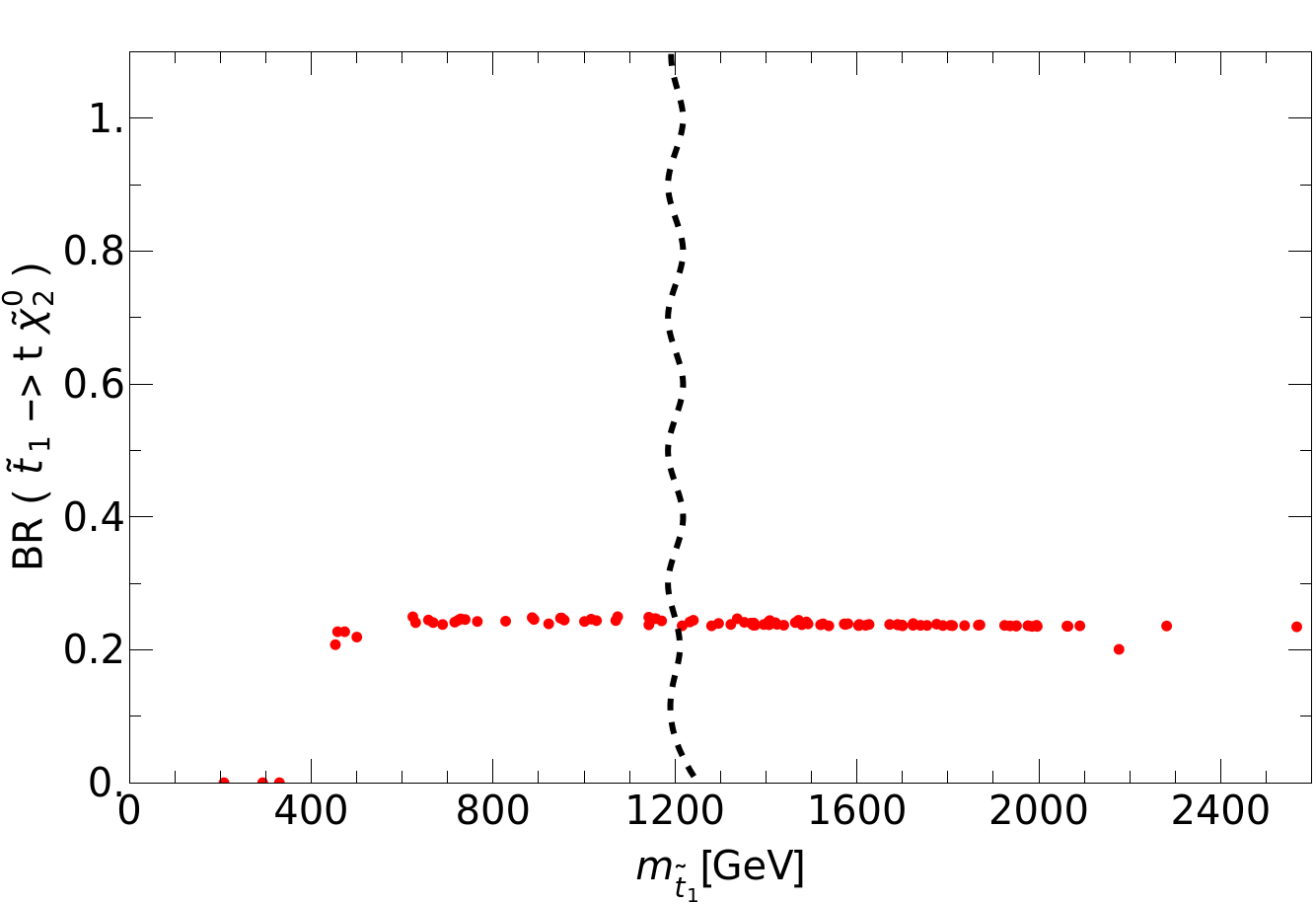}
    \caption{The variation of the branching ratios of $\widetilde{t}_1$ vs. the stop mass, $m_{\widetilde{t_1}}$ into the different decay modes.}
    \label{fig:br}
\end{figure}
\begin{figure}
\centering{\includegraphics[scale=0.30]{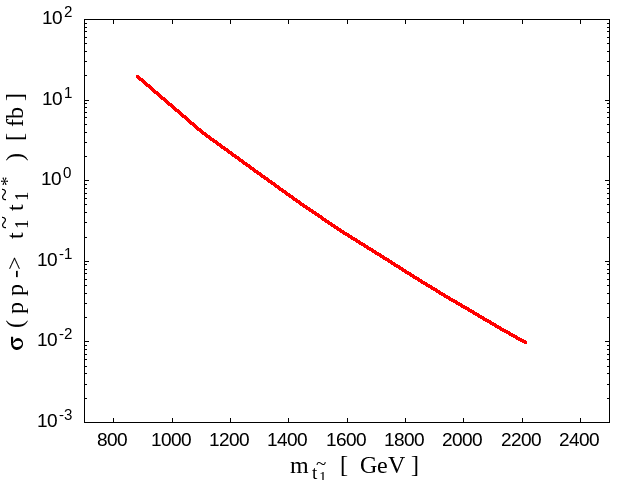}}
\caption{Stop pair production cross-section at NLO at $\sqrt{s}=14$ TeV at the HL-LHC.}
\label{fig:stopcs}
\end{figure}
\section{Collider study}  
\begin{figure}[ht]
    \centering
    \includegraphics[scale=0.08]{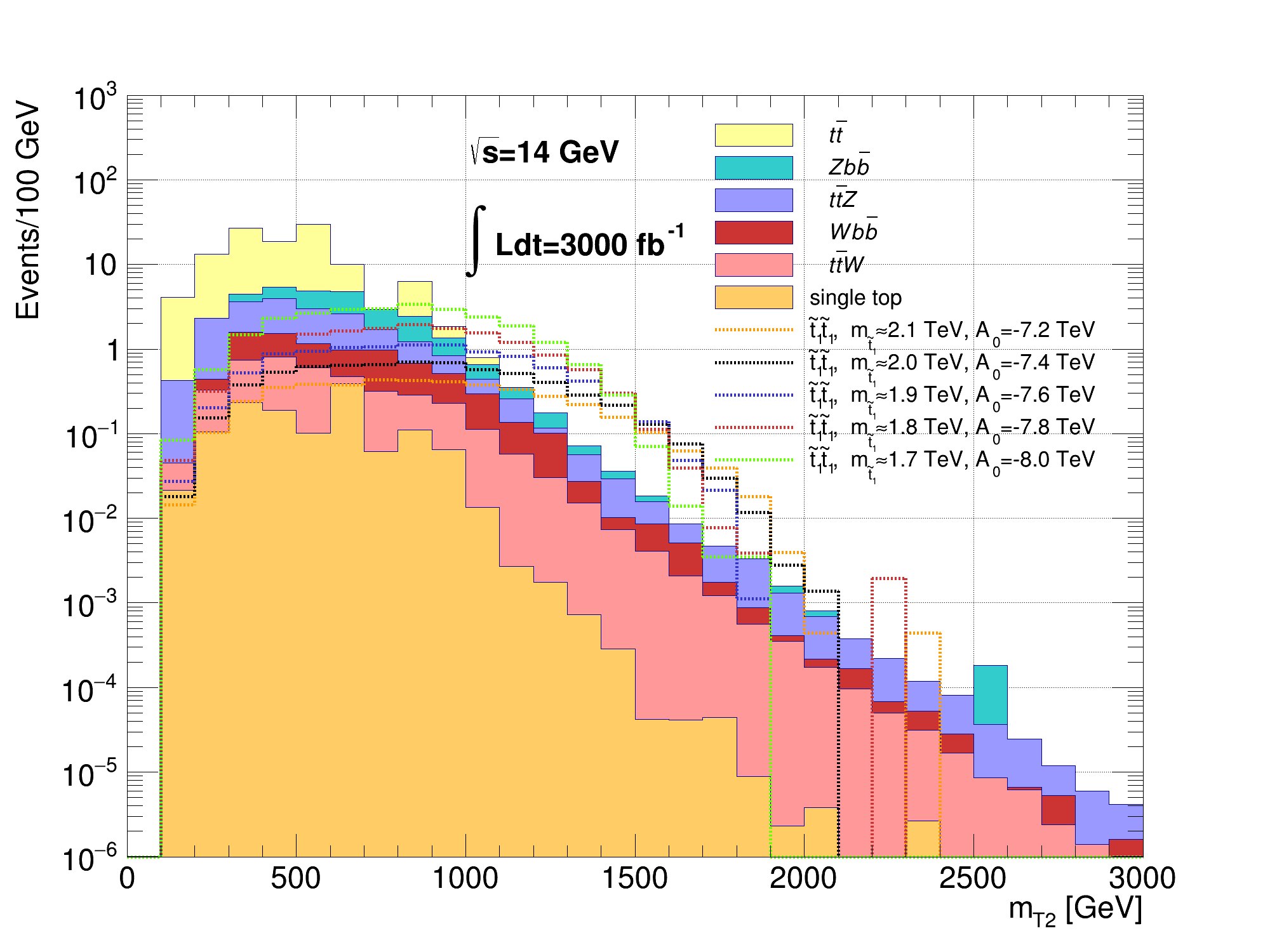}
       \includegraphics[scale=0.32]{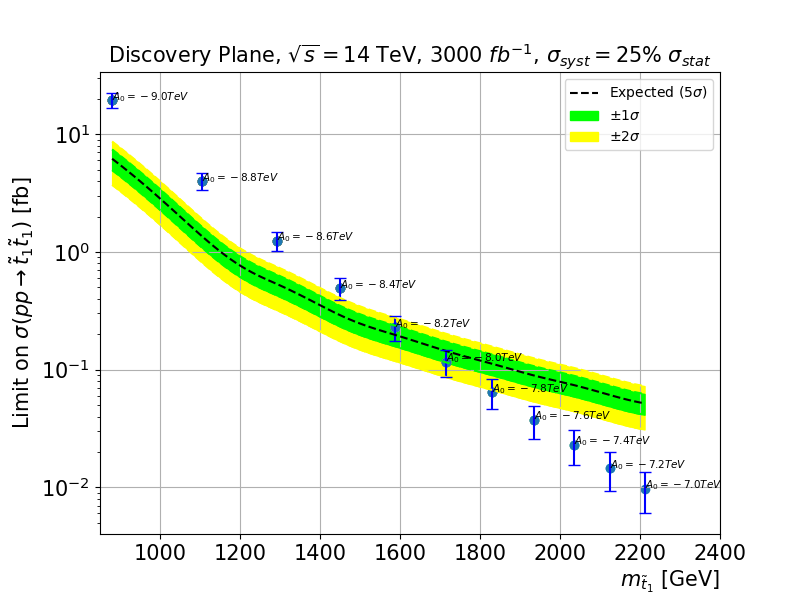} 
     \caption{Distribution of $m_{T_2}$ for the final state 1$b$+1$t$ + $\slashed{E}_T$ and expected 5$\sigma$ discovery plane for the lightest stop at the HL-LHC.}
    \label{fig:mt2}
\end{figure}

In this section, we discuss the collider prospects of the lightest stops at the high luminosity LHC (HL-LHC).  We choose a benchmark point shown in Table ~\ref{tab:bm} consistent with current experimental constraints from flavour physics and LHC data and scan over the parameter $A_t$ to vary the mass of the stop in the range 800-2200 GeV. The production cross-section of the lightest stops is shown in Fig.~\ref{fig:stopcs} for $\sqrt{s}=14$ TeV. For the collider analyses, we consider the signal final states: $b\bar{b}+\slashed{E}_T$, 1 t + 1 b $+ \slashed{E}_T$ and 2 t $+ \slashed{E}_T$.   The dominant SM backgrounds are $t\bar{t}, b\bar{b}Z, t\bar{t}Z, t\bar{t}W, b\bar{b}W$ and single top production channel. 
Using boosted jet techniques to reconstruct top jets with $p_T>400$ GeV and $R=1.5$ and relying on hard kinematic observables such as $\slashed{E}_T>400$ GeV, $H_T>1400 $ GeV, $L_T(=H_T+\slashed{E}_T)>1400$ GeV, min($m_T(t,\slashed{E}_T),m_T(b,\slashed{E}_T))\geq175$ GeV, $\Delta \Phi ( b, \slashed{E}_T)\geq40^{\circ} $, $\Delta \Phi ( J, \slashed{E}_T)\geq30 ^{\circ}$ to suppress the SM background for the $tb+\slashed{E}_T$, we observe that the key kinematic variable to discriminate between signal and backgrounds is $M_{T_2}$ especially at the tails of the distribution as shown in Fig.~\ref{fig:mt2} for $\sqrt{s}=14 $ TeV and integrated luminosity of $3000$ fb$^{-1}$. Similar cuts (see ~\cite{Baer:2023uwo}) are imposed on the $bb+\slashed{E}_T$ and $tt + \slashed{E}_T$ channels. In all cases, the top-squark pair production is revealed as an enhancement in the $m_{T_2}$ distribution at high values of $m_{T_2}$. 
A combined reach of all channels at HL-LHC lead to a reach of $\sim$1.7 TeV at 5$\sigma$  as seen in Fig.~\ref{fig:mt2}(right) and $\sim$2 TeV  at 2$\sigma$~\cite{Baer:2023uwo}. 
\begin{table}[h!]
 \scriptsize
\centering
\begin{tabular}{lc}
\hline
Parameters &  Benchmark point \\
\hline
$m_0$      & 5 TeV \\
$m_{1/2}$      & 1.2 TeV \\
$A_0$      & -8 TeV \\
$\tan\beta$    & 10  \\
\hline
$\mu$  & 250 GeV  \\
$m_A$  & 2 TeV \\
\hline
$m_{\tilde{g}}$   & 2830 GeV \\
$m_{\tilde{t}_1}$& 1714 GeV \\
$m_{\tilde{t}_2}$& 3915 GeV \\
$m_{\tilde{\chi}_1^\pm}$ & 261.7 GeV \\
$m_{\tilde{\chi}_2^\pm}$ & 1020.6 GeV \\
$m_{\tilde{\chi}_1^0}$ & 248.1 GeV \\ 
$m_{\tilde{\chi}_2^0}$ & 259.2 GeV \\ 
$m_{\tilde{\chi}_3^0}$ & 541.0 GeV \\ 
$m_{\tilde{\chi}_4^0}$ & 1033.9 GeV \\ 
$m_h$       & 124.7 GeV \\ 
\hline
$\Omega_{\tilde{\chi}_1}^{std}h^2$ & 0.016 \\
$\sigma^{SI}(\tilde{\chi}_1^0, p)$ (pb) & $2.2\times 10^{-9}$ \\
$\sigma^{SD}(\tilde{\chi}_1^0, p)$ (pb)  & $2.9\times 10^{-5}$ \\
$\Delta_{\rm EW}$ & 22 \\
\hline
\end{tabular}
\caption{Input parameters (TeV) and masses (GeV) for the stringy natural SUSY benchmark point from the NUHM2 model with $m_t=173.2$ GeV using Isajet 7.88.}
\label{tab:bm}
\end{table} 
\section{Summary and Conclusions}
In this work, we have investigated the properties of  top squarks from the string landscape where a power-law draw to large soft terms is expected. The derived value of the weak scale must lie within the ABDS window in order to allow for complex nuclei (and hence atoms) in each anthropically-allowed pocket universe. Under this stringy naturalness requirement, we find $m_{\widetilde{t}_1}\sim$ 1-2.5 TeV with large mixing which also facilitates to lift $m_h$ to 125 GeV while minimizing the top squark contributions to the weak scale. Despite of the large mixing, the lighter top-squark is mainly a right-squark,  and lead to mixed final states of $b\bar{b}+ \slashed{E}_T$, $t\bar{t}+ \slashed{E}_T$ and $tb+ \slashed{E}_T$.  Using boosted jet techniques to investigate the reach of stops at HL-LHC  it is possible to reach $m_{\widetilde{t}_1}\simeq 1.7$ at 5$\sigma$ and  $\simeq 2$ TeV at 2$\sigma$ therefore covering most of the stringy natural parameter space at HL-LHC. 
\section*{Acknowledgments}
The author acknowledges support from the  HEP  Dodge Family Endowment Fellowship at the Homer L.Dodge Department of Physics $\&$ Astronomy at the University of Oklahoma. The author thanks her collaborators H.Baer, V.Barger, D.Sengupta and K.Zhang for the successful completion of the work Ref.~\cite{Baer:2023uwo}, on which this talk is based on.


\begin{thebibliography}{99}
\bibitem{Barbieri:1987fn}
R.~Barbieri and G.~F.~Giudice,
Nucl. Phys. B \textbf{306} (1988), 63-76
doi:10.1016/0550-3213(88)90171-X
\bibitem{Baer:2013gva}
H.~Baer, V.~Barger and D.~Mickelson,
Phys. Rev. D \textbf{88} (2013) no.9, 095013
doi:10.1103/PhysRevD.88.095013
[arXiv:1309.2984 [hep-ph]].

\bibitem{Douglas:2012bu}
M.~R.~Douglas,
doi:10.1142/9789814412551\_0012
[arXiv:1204.6626 [hep-th]].
\bibitem{Agrawal:1997gf}
V.~Agrawal, S.~M.~Barr, J.~F.~Donoghue and D.~Seckel,
Phys. Rev. D \textbf{57} (1998), 5480-5492
doi:10.1103/PhysRevD.57.5480
[arXiv:hep-ph/9707380 [hep-ph]].
\bibitem{Baer:2023uwo}
H.~Baer, V.~Barger, J.~Dutta, D.~Sengupta and K.~Zhang,
doi:10.1103/PhysRevD.108.075027
[arXiv:2307.08067 [hep-ph]].
\bibitem{Baer:2012up}
H.~Baer, V.~Barger, P.~Huang, A.~Mustafayev and X.~Tata,
Phys. Rev. Lett. \textbf{109} (2012), 161802
doi:10.1103/PhysRevLett.109.161802
[arXiv:1207.3343 [hep-ph]].
\bibitem{Weinberg:1987dv}
S.~Weinberg,
Phys. Rev. Lett. \textbf{59} (1987), 2607
doi:10.1103/PhysRevLett.59.2607
\bibitem{Baer:2023cvi}
H.~Baer, V.~Barger, D.~Martinez and S.~Salam,
Phys. Rev. D \textbf{108} (2023) no.3, 035050
doi:10.1103/PhysRevD.108.035050
[arXiv:2305.16125 [hep-ph]].
\bibitem{Brust:2011tb}
C.~Brust, A.~Katz, S.~Lawrence and R.~Sundrum,
JHEP \textbf{03} (2012), 103
doi:10.1007/JHEP03(2012)103
[arXiv:1110.6670 [hep-ph]].
\bibitem{Baer:2020kwz}
H.~Baer, V.~Barger, S.~Salam, D.~Sengupta and K.~Sinha,
Eur. Phys. J. ST \textbf{229} (2020) no.21, 3085-3141
doi:10.1140/epjst/e2020-000020-x
[arXiv:2002.03013 [hep-ph]].

\bibitem{Brust:2011tb}
C.~Brust, A.~Katz, S.~Lawrence and R.~Sundrum,
JHEP \textbf{03} (2012), 103
doi:10.1007/JHEP03(2012)103
[arXiv:1110.6670 [hep-ph]].
\bibitem{Kitano:2006gv}
R.~Kitano and Y.~Nomura,
Phys. Rev. D \textbf{73} (2006), 095004
doi:10.1103/PhysRevD.73.095004
[arXiv:hep-ph/0602096 [hep-ph]].
\bibitem{Baer:2013gva}
H.~Baer, V.~Barger and D.~Mickelson,
Phys. Rev. D \textbf{88} (2013) no.9, 095013
doi:10.1103/PhysRevD.88.095013
[arXiv:1309.2984 [hep-ph]].
\bibitem{Baer:2023cvi}
H.~Baer, V.~Barger, D.~Martinez and S.~Salam,
Phys. Rev. D \textbf{108} (2023) no.3, 035050
doi:10.1103/PhysRevD.108.035050
[arXiv:2305.16125 [hep-ph]].
\bibitem{Mustafayev:2014lqa}
A.~Mustafayev and X.~Tata,
Indian J. Phys. \textbf{88} (2014), 991-1004
doi:10.1007/s12648-014-0504-8
[arXiv:1404.1386 [hep-ph]].
\bibitem{Baer:2014ica}
H.~Baer, V.~Barger, D.~Mickelson and M.~Padeffke-Kirkland,
Phys. Rev. D \textbf{89} (2014) no.11, 115019
doi:10.1103/PhysRevD.89.115019
[arXiv:1404.2277 [hep-ph]].

\end{thebibliography}
\end{document}